# Deuterium – Tritium pulse propulsion with hydrogen as propellant and the entire space-craft as a gigavolt capacitor for ignition.

By F. Winterberg
University of Nevada, Reno

**Abstract**

A deuterium-tritium (DT) nuclear pulse propulsion concept for fast interplanetary transport is proposed utilizing almost all the energy for thrust and without the need for a large radiator:

1. By letting the thermonuclear micro-explosion take place in the center of a liquid hydrogen sphere with the radius of the sphere large enough to slow down and absorb the neutrons of the DT fusion reaction, heating the hydrogen to a fully ionized plasma at a temperature of ~ $10^5$ K.

2. By using the entire spacecraft as a magnetically insulated gigavolt capacitor, igniting the DT micro-explosion with an intense GeV ion beam discharging the gigavolt capacitor, possible if the space craft has the topology of a torus.

# 1. Introduction

The idea to use the 80% of the neutron energy released in the DT fusion reaction for nuclear micro-bomb rocket propulsion, by surrounding the micro-explosion with a thick layer of liquid hydrogen heated up to $10^5$ K thereby becoming part of the exhaust, was first proposed by the author in 1971 [1]. Unlike the Orion pusher plate concept, the fire ball of the fully ionized hydrogen plasma can here be reflected by a magnetic mirror.

The 80% of the energy released into 14MeV neutrons cannot be reflected by a magnetic mirror for thermonuclear micro-bomb propulsion. This was the reason why for the Project Daedalus interstellar probe study of the British Interplanetary Society [2], the neutron poor deuterium-helium 3 ($DHe^3$) reaction was chosen. For an interstellar mission the highest possible specific impulse is required, but one also wants to avoid a large radiator to remove the heat released by neutrons absorbed in the space craft. Because of the rarity of the helium 3 isotope, it was proposed to mine $He^3$ from the atmosphere of Jupiter. But for a manned mission inside the solar system, in particular to the planet Mars, no such high specific impulse is needed or even desired. There an exhaust velocity of 30 km/s is sufficient, whereby a trip to Mars would be as short as one week. This can be achieved by heating a hydrogen propellant to ~$10^5$ K.

Besides the Daedalus study, only two other studies went into engineering details: The VISTA study of the Lawrence Livermore National Laboratory [3], and the somewhat related ICAN concept by Pennsylvania State University [4].

While the Daedalus concept proposes to use intense relativistic electron beams for the ignition of the thermonuclear micro explosions (as it was proposed by author [1]), the VISTA concept proposes to use lasers, with the ICAN concept [4] using antiprotons for ignition, as "the odd man out".

All these concepts have their advantages and disadvantages. In favor of electron beams speaks the high efficiency by which they can be generated, but they are difficult to focus and to stop over a short distance, both required for ignition. For laser beams the situation is reversed. In general high efficiency is needed if one wants to avoid a large radiator to remove the waste heat from the space craft. More serious is the heating of the space craft by the absorption of neutrons, in particular for the DT reaction where 80% of the energy goes into neutrons. But even the $DHe^3$ reaction is not completely without the release of some neutrons, through the unavoidable DD reactions in a hot plasma. While for interstellar missions the highest possible specific impulse is mandated, this problem is less serious for an interplanetary missions, because can there surround the DT micro-explosion with a thick neutron-absorbing layer of liquid hydrogen, to be heated up by the neutrons to a high temperature plasma of ~ $10^5$ K and deflected by a magnetic mirror.

The stopping length of the neutrons is determined by the material and the density of this material, but the mass of the material stopping the neutrons is determined by its geometric arrangement. With the thermonuclear energy released by the almost point-like source of a thermonuclear micro-explosion, the smallest mass is realized if the micro-explosion occurs in the center of a neutron-stopping sphere, and to reach the highest possible temperature the sphere should be of liquid hydrogen, or even better pre-compressed liquid hydrogen. Pre-compression could be achieved by surrounding the liquid hydrogen sphere with a layer of high explosive, simultaneously ignited on its outer surface. Under a ten-fold compression, the radius of the hydrogen sphere would be reduced by the factor $10^{1/3}$ ~2.15, with the amount of hydrogen reduced ten-fold.

It is here proposed to place the DT thermonuclear target in the center of liquid hydrogen sphere, with the target to be ignited by a GeV ion beam—passing through a pipe (see Fig. 1). To increase energy output, the hydrogen sphere can be surrounded by a shell made from a neutron absorbing boron. The energy released as energetic α-particles by the absorption of the neutrons in the boron not only increases the overall energy output, but also compresses the hydrogen sphere.

Following the ignition and burn of the DT target, the hydrogen is converted into an expanding hot plasma fire ball used for rocket propulsion.

For this idea to work, the radius of the liquid hydrogen sphere must be large enough to slow down and stop the neutrons, but not be larger than is required to keep its temperature at about $10^5$ K. This condition can be met for liquid hydrogen spheres of reasonable dimensions.

## 2. The Neutron Physics of the Proposed Scheme

As in fission reactors, the neutron physics is determined by the slowing down and diffusion of the neutrons in the blanket. Assuming that radius of the DT target is small compared to the outer radius of the neutron-absorbing blanket, one can approximate the neutron source of the burning DT target as a point source.

We use the Fermi age theory [2] for the slowing down of the neutrons from their initial energy $E_o$ to their final energy E. The neutron-slowing down is determined by the Fermi age equation:

$$\frac{\partial q}{\partial \tau} = \nabla^2 q \tag{1}$$

where the "age" $\tau$ is given by:

$$\tau(E) = \int_E^{E_0} \frac{D}{\xi \Sigma_s E} dE \tag{2}$$

D is the neutron diffusion constant,

$$D = \frac{1}{3\Sigma_s (1-\mu_0)} \tag{3}$$

with $\Sigma_s$ the macroscopic scattering cross section, $\Sigma_s = n\sigma_S$, where n is the particle number density in the blanket, and $\sigma_s$ is the scattering cross section. Furthermore,

$$\mu_0 = \frac{2}{3A} \tag{4}$$

is a scattering coefficient for a substance of atomic weight A. For hydrogen, one has A = 1 and hence $\mu_o = 2/3$, making $D = 1/\Sigma_s$. The logarithmic energy decrement of the neutron deceleration is given by:

$$\xi = 1 + \frac{(A-1)^2}{2A} \ln \frac{A-1}{A+1} \tag{5}$$

For A = 1, one has $\xi = 1$.

Setting $E_0$ = 14MeV, E = 10eV ($10^5$K), and using the neutron physics data of the Brookhaven National Laboratory [3], one finds that $\tau \cong 6 \times 10^2$ cm$^2$. This implies a slowing down length of the order $\sqrt{\tau} \cong 20$ cm. For water, one has, by comparison for the $E_0$=2MeV fission energy, neutrons slowed down to the thermal energy E = 2 × $10^{-2}$eV, $\tau$ = 33cm$^2$, and $\sqrt{\tau}$ = 5.7cm. However, for the production of a $10^5$K fire ball, water is unsuitable because at these temperatures most of the energy goes into blackbody radiation. For this reason alone, hydrogen is to be preferred.

The expression for $\tau$ given by (2) ignores neutron absorption during the slowing down process. If taken into account, one has to multiply $\tau$(E) with the resonance escape probability p(E) given by

$$p(E) = \exp\left(-\frac{1}{\xi} \int_E^{Eo} \frac{\Sigma_a}{\Sigma_a + \Sigma_s} \cdot \frac{dE}{E}\right) \tag{6}$$

where $\Sigma_a$ = $n_a\sigma_a$ is the macroscopic absorption cross-section, with $n_a$ as the particle number density of the neutron absorbing substance, and $\sigma_a$ is the microscopic absorption cross-section. For graphite-moderated reactors p ≈ 0.5. For large $\sigma_a$ and even if $n_a$ << n, p(E) can substantially reduce $\tau$ and hence the stopping length.

Another important number is the slowing down time for the neutrons, given by

$$t_0 = \frac{\sqrt{2M}}{\xi\overline{\Sigma}_s}\left(\frac{1}{\sqrt{E_{th}}} - \frac{1}{\sqrt{E_0}}\right) \tag{7}$$

where M is the neutron mass.

For liquid hydrogen $t_0 \approx 10^{-5}$ sec. This time is much longer than the time of the DT micro-explosion, but it must be about equal to the inertial expansion time of the fire ball with an initial radius R. For an expanding plasma fire ball of initial radius R and expansion velocity V,

$$t_0 \cong \frac{R}{V} \tag{8}$$

At a temperature of $10^5$ K, the expansion velocity is V $\approx$ 30 km/s and setting $R \cong \sqrt{\tau} = 20$ cm, one has $t_o \cong 10^{-5}$ sec. For liquid hydrogen where n = $5 \times 10^{22}$ cm$^{-3}$, the total number of hydrogen atoms in a spherical volume with a radius of 20 cm, is of the order N = $2 \times 10^{27}$. Heated to a temperature of T = $10^5$ K, the thermal energy of the fire ball is of the order E $\cong$ NkT $\approx 3 \times 10^{16}$ erg = 1 ton of TNT. For a DT target requiring an ignition energy of 1-10MJ, this is a gain of the order 300.

## 3. Spatial Distribution of the Decelerated Neutrons

Making the point-source approximation for the neutrons released from the DT fusion micro-explosion, their spatial distribution by the Fermi age equation is

$$q(r,t) = \frac{e^{-r^2/4\tau}}{(4\pi\tau)^{\frac{3}{2}}} \qquad (9)$$

The slowing down of the neutrons is followed by their diffusion which is ruled by the diffusion equation in spherical coordinates

$$D \frac{1}{r^2} \frac{\partial}{\partial r}\left(r^2 \frac{\partial \phi}{\partial r}\right) - \Sigma_a \phi + S = \frac{\partial n}{\partial t} \qquad (10)$$

where $\phi$ is the neutron flux, $\Sigma_a = n\sigma_a$, with $\sigma_a$ the neutron absorption cross-section. S is the neutron source given by pq.

Surrounding the hydrogen by boron, the diffusion equation must be solved with the boundary condition for the neutron flux in the hydrogen A, and boron B

$$\left. \begin{array}{c} \phi_A = \phi_B \\ D_A \dfrac{d\phi_A}{dr} = D_B \dfrac{d\phi_B}{dr} \end{array} \right\} \qquad (11)$$

Because of the very large neutron absorption cross section for thermal (or epithermal) neutrons, only a comparatively thin layer of boron is needed. The 3MeV of energy released in charged particles by the neutron absorption of neutrons in boron, can be simply added to the 14MeV of the neutron energy of the DT reaction.

## 4. Propulsion

The way propulsion is achieved is straightforward and explained in Fig. 2. The mini-fusion bomb F, is catapulted into the focus of a parabolic magnetic reflector R, made from steel and placed inside a large magnetic field coil C.

The inner side of the reflector is insulated against the hot plasma of the expanding fire ball by the thermomagnetic Nernst effect. It generates currents in the boundary layer between the cool wall and the hot plasma of the fire ball, producing a magnetic field between the wall and the fire ball just sufficiently strong to repel the fire ball from the wall [8].

For a temperature of $T \approx 10^5$ K, the hydrogen is a fully ionized plasma and can be deflected by a magnetic mirror [1]. At a temperature of $T \approx 10^5$ K the exhaust velocity is about equal to the thermal expansion velocity of a hydrogen plasma which at this temperature is $V \cong 30$ km/s. Assuming a mass ratio equal to 10, this implies a final speed of the space craft of the order 100 km/s. With this velocity, and assuming the smallest distance of the earth from Mars of about 50 million km, the travel time is of the order $5 \times 10^5$ sec, or about one week. The velocity of 100 km/s must be reached in a time less than the travel time, for example in $10^5$ sec. With the thrust given by $T = vdm/dt$, assuming for the exhaust $dm/dt = 0.1$ ton/sec, one has $T = 3 \times 10^{11}$ dyn. Assuming an acceleration of the space craft equal to $g \cong 10^3$ cm/s$^2$, the mass of the space craft then is $M = T/g = 10^8$ g = 100 tons, large enough for an unmanned space craft, but it can be easily scaled up to 1000 tons, as required for a manned mission.

The ignition of the DT thermonuclear micro-explosion can be performed by a $10^7$ Ampere-GeV proton beam, generated by using the entire space craft as a large, magnetically insulated capacitor, inductively charged up to gigavolt potentials, leading to an electron cloud in the vacuum surrounding the space craft [5]. For the injection of the GeV proton beam into the pipe reaching the DT target as shown in Fig.1, the liquid hydrogen sphere must be grounded against the electron cloud surrounding the space craft. This can be done by a small plasma jet emitted from the surface of the hydrogen sphere as shown in Fig. 3. The jet can be produced by a laser beam heating the material making up the jet.

The recharging of the entire space craft to gigavolt potentials by inductive charge injection is shown in Fig. 2. The electromagnetic pulse induced in the induction loop surrounding the rocket propulsion chamber, is used to radially inject electrons at the top of the space craft into the rising magnetic field of an auxiliary coil. Apart from a different design of the rocket propulsion chamber, this is the same kind of inductive charge injection for a pure deuterium bomb propulsion [5].

The idea to use a large magnetically insulated and levitated torus, to be charged up to gigavolt potentials, was already proposed by the author in 1968 [9,10]. This configuration is topologically the same as the one used for the space craft shown in Fig.2.

## 6. Conclusion

Because it uses the entire space craft as a gigavolt capacitor, to be discharged into an intense ion beam for ignition, the proposed concept is expected to be superior to all other designs intended to reach Mars in the shortest possible time. All the other designs depend on a massive radiator. To reduce the mass of the radiator, "droplet" radiators have been proposed, but they suffer from the loss of droplet mass by evaporation. Not only the VISTA and ICAN concept, but also various electric propulsion concepts cannot compete with the compact gigavolt capacitor concept ideally suited for the high vacuum of space.

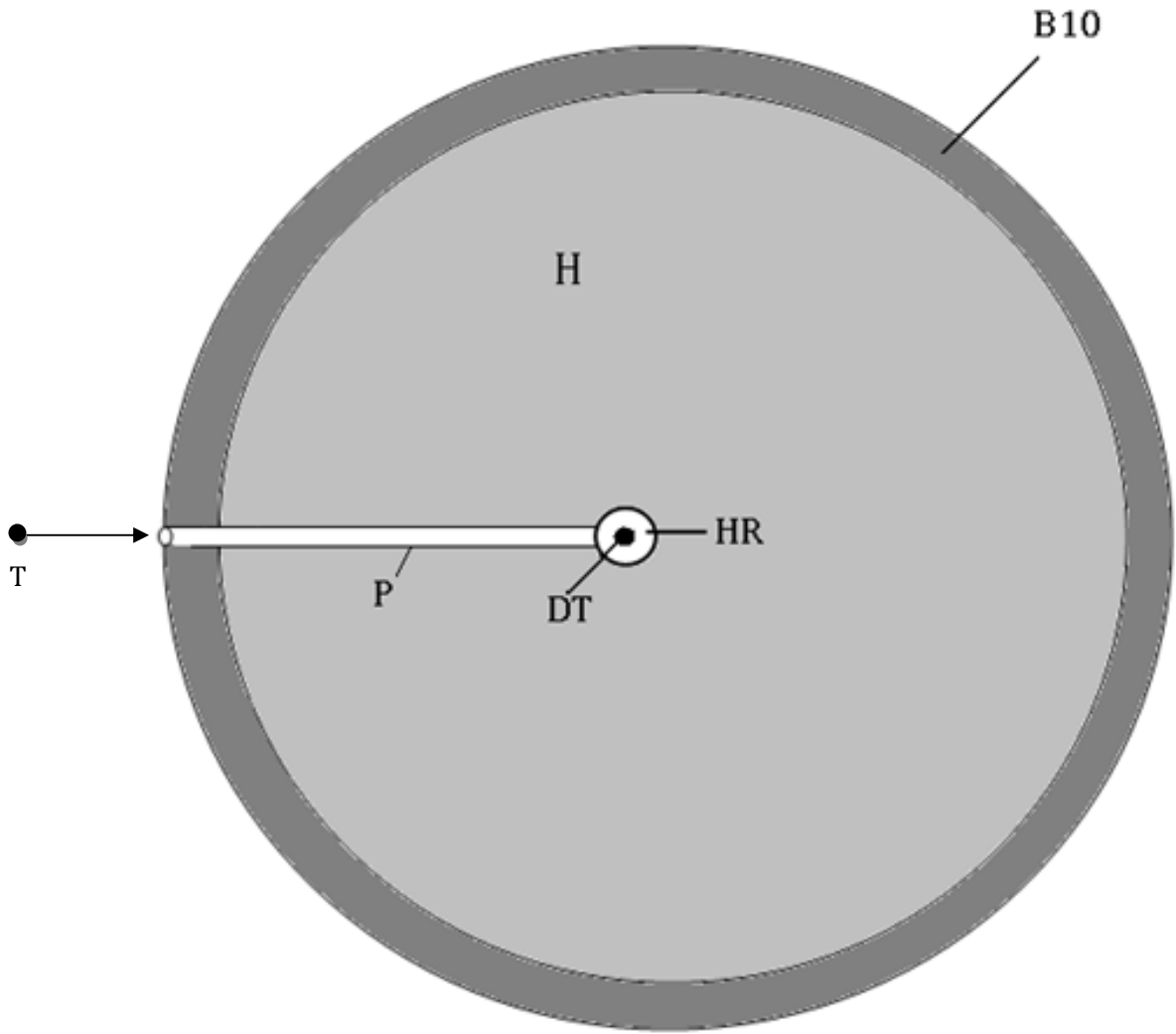

**Figure 1**

DT deuterium-tritium fusion target T ignited by a GeV ion beam, with the target B catapulted through thin pipe P into the center of the sphere. H liquid hydrogen, B10 solid boron.

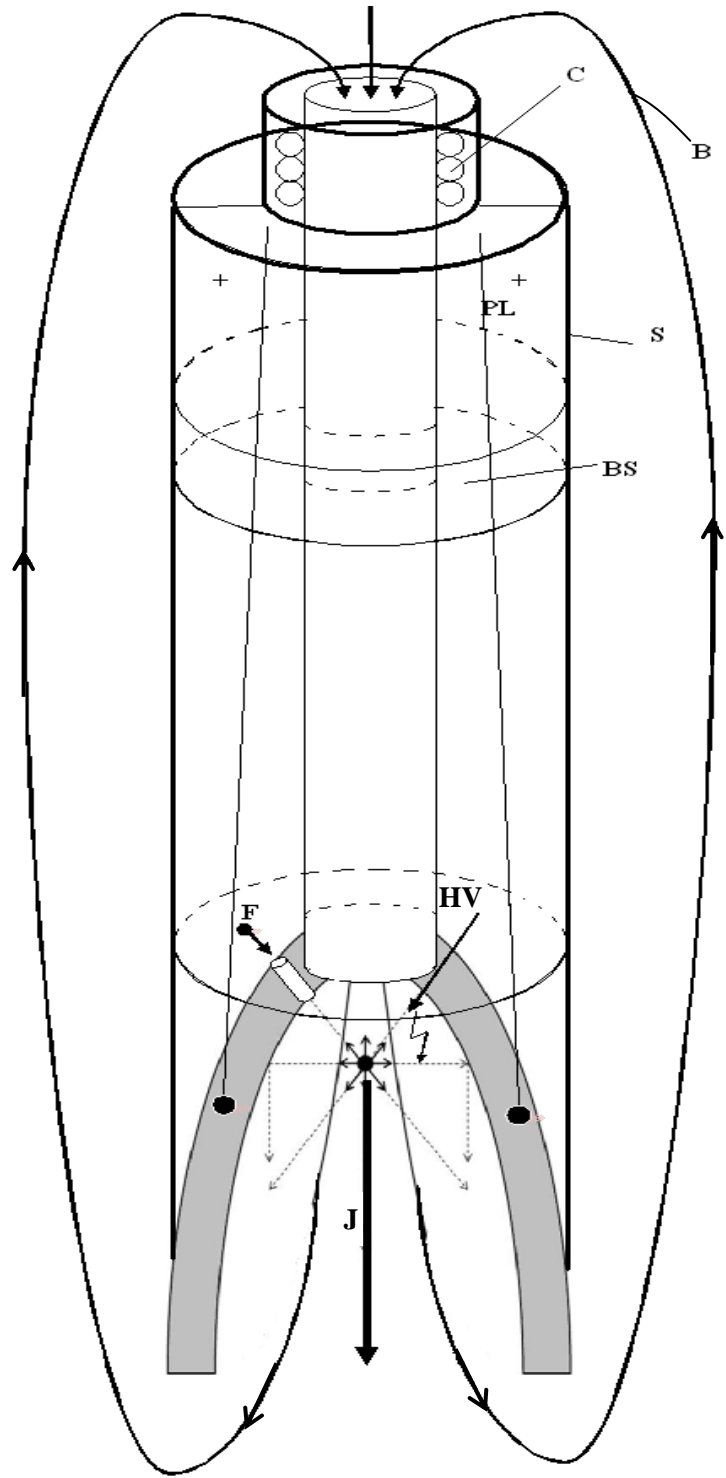

**Figure 2**
**DT pulse propulsion rocket, with spacecraft acting as large toroidal gigavolt capacitor to ignite thermonuclear micro explosion; F mini-fusion bomb assembly shot into focus of magnetic mirror, J plasma jet; HV GeV proton beam; B magnetic mirror field; BS biological shield; PL payload.**

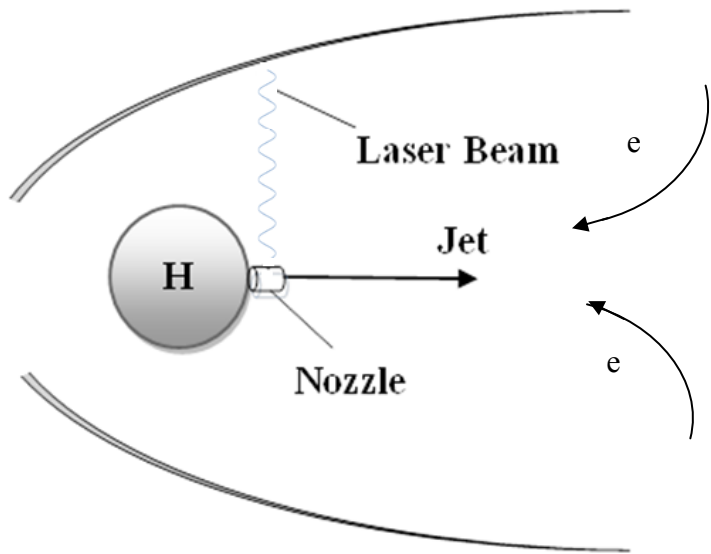

**Figure 3**
**Grounding of the hydrogen sphere against the electron cloud surrounding the space-craft by a small laser driven jet; e electrons**